# Mutual phase-locking in high frequency microwave nanooscillators as function of field angle

**G. Hrkac**[1], T. Schrefl[1], S. Bance[1], D. Allwood[1], A. Goncharov[1], J. Dean[1]

D. Suess[2]

[1] *Department of Engineering Materials, University of Sheffield, Sheffield, UK.*

[2]*Solid State Physics, Vienna University of Technology, Vienna, Austria*

Abstract

We perform a qualitative analysis of phase locking in a double point-contact spinvalve system by solving the Landau-Lifshitz-Gilbert-Slonzewski equation using a hybrid-finite-element method. We show that the phase-locking behaviour depends on the applied field angle. Starting from a low field angle, the locking-current difference between the current through contact A and B increases with increasing angle up to a maximum of 14 mA at 30° and it decreases thereafter until it reaches a minimum of 1 mA at 75°. The tunability of the phase-lock frequency with current decreases linearly with increasing out of plane angle from 45 to 21 MHz/mA.

It has been theoretically predicted [1-3] and experimentally observed [4-6] that spin-polarized currents in magnetic nanostructure devices may excite steady state microwave magnetization precessions in ferromagnetic thin films. Recent experiments have demonstrated that microwave oscillations in current driven nanocontact devices can be phase-locked in frequency by varying the bias current through one of the contacts [7,8].

A lot of work has focused on understanding the interaction of spin-polarized current and ferromagnetic nanostructures [9, 10, 11] and several models and studies, analytical and numerical [12, 13, 14, 15], have been presented to explain experimental results. It is crucial for the design of magneto-electronic nano-oscillators to develop reliable simulation techniques in order to predict relevant parameters such as the threshold current to excite limit-cycles of oscillatory motion of the magnetization, the amplitude and the current dependence of the limit-cycle frequency as well as its dependence on external or biased field sources [5].

We have developed a three dimensional micromagnetic model to describe spin-torque effects in point contact geometries. In particular we investigate the mutual phase locking of current-driven oscillations between two point contacts [8]. Phase locking was shown experimentally in double point contacts for an applied external field at field angles of 75 degree and above [8]. Interestingly phase locking only occurs when there is a difference in the current between the two point contacts. At an out of plane field angle of 75 degrees the locking current difference is 1 mA. In this letter we investigate how the locking current difference depends on the applied field angle. For the double point contact regime we apply the field where the projection of the field onto the film coincides with the line connecting the two point contacts. The field angles are varies from 15 to 75 degrees.

The letter is organized as follows. First we describe the micromagnetic model and validate the model by comparing simulations results with experimental data by Rippard and co-workers [5]. We solve the Landau-Lifshitz-Gilbert equation expanded with the asymmetric Slonczewski spin torque term [1] and the quasi-static Maxwell equations. The model takes into account the inhomogeneous current flow and the associated Oersted field. The local current density through the device is calculated dynamically as function of the magnetization distribution whereby the magnetoresistivity is allowed to change locally depending on the angle between the magnetization in the free layer and in the reference layer. Further we use a new boundary method to minimize the reflection from the boundaries based on scattering theory [20].

Next we characterize the conditions for mutual phase locking using micromagnetic simulations. In the phase locking regime the power spectrum of the film's magnetization shows a dominating single peak. The power of this coherent mode is over twice that of the individual oscillations without phase locking. Phase locking occurs for a certain locking current difference, $\Delta I_{AB}$, the difference between the currents through contact A and contact B. The common frequency can be tuned in up to maximum range, $\Delta f_{max}$, by changing the current. Both, $\Delta I_{AB}$ and $\Delta f_{max}$ depend on the out of plane angle at which the field is applied.

The micromagnetic description of the microwave generation in the nanocontacts start from the total magnetic Gibbs free energy $E_{tot}$ [16]

$$E_{tot} = \int_V \left( \frac{A}{M_s^2} (\nabla \mathbf{M})^2 + E_{anis} - \mu_0 \mathbf{M} \left( \mathbf{H}_a + \frac{\mathbf{H}_d}{2} \right) \right) dV \quad (1)$$

which is discretized using a finite element method that is solved using a hybrid finite element/boundary element method [17, 18]. The first term in equation 1 is the exchange energy followed by the anisotropy energy density, the applied field $\mathbf{H}_a$ and the magnetostatic energy with $\mathbf{H}_d$ being the demagnetizing field, with $\mu_0$ being the permeability of free space. The nonlinear magnetization dynamics of the nanocontact is described by a modified Landau-Lifshitz-Gilbert equations, which is written in the following dimensionless form

$$(1+\alpha^2)\frac{d\mathbf{m}}{d\tau} = -\mathbf{m} \times \mathbf{h}_{\text{eff}} - \alpha \mathbf{m} \times (\mathbf{m} \times \mathbf{h}_{\text{eff}}) - \mathbf{N} \qquad (2)$$

with the spin transfer torque $\mathbf{N}$ describing the effects caused by the spin polarized current. The magnetization vector $\mathbf{m} = \mathbf{M}/M_s$ is the magnetization $\mathbf{M}$ normalized by the saturation magnetization $M_s$, and is assumed to be spatially nonuniform, $\alpha$ is the Gilbert damping constant, $\tau$ the time measured in units of $(|\gamma| M_s)^{-1}$ ($\gamma$ is the gyromagnetic ratio) and $\mathbf{h}_{\text{eff}} = \mathbf{H}_{\text{eff}}/M_s$ is the normalized effective field. The effective field is the negative variational derivative of the total magnetic Gibbs free energy density (1). We perform zero temperature simulations and do not include a random thermal field in the effective field. The spin transfer torque $\mathbf{N}$ in the asymmetric form [19] can be written as

$$\mathbf{N} = \frac{\hbar J_e}{2ed} \frac{1}{\mu_0 M_s^2} g_T \mathbf{m} \times (\mathbf{m} \times \mathbf{p}) = \frac{1}{2} \frac{J_e}{J_p} g_T \mathbf{m} \times (\mathbf{m} \times \mathbf{m}_p) \qquad (3)$$

Here $J_e$ is the electron current density, $J_p = (\mu_0 e d M_s^2 / \hbar)$ the characteristic current density of the system ($\hbar$ is the Planck constant, $e$ the electron charge, $d$ the free layer thickness) and $\mathbf{m}_p$ is the unit vector in the direction of the magnetization in the reference layer. For a NiFe free layer with a thickness of 5 nm and a saturation magnetization of $7.16 \times 10^5$ A m$^{-1}$ one obtains a characteristic current density $J_p$ of $6.53 \times 10^{13}$ A m$^{-2}$. The scalar function $g_T$ is defined as follows

$$g_T(\theta) = \left( \frac{8P^{3/2}}{3(1+P)^3 - 16P^{3/2} + (1+P)^3 \cos(\theta)} \right) \qquad (4)$$

Where $P$ is the spin polarization factor and $\cos(\theta)=\mathbf{m}\cdot\mathbf{m_P}$ is the angular function of the normalized magnetization of the free layer $\mathbf{m}$ and the fixed layer $\mathbf{m_P}$.

The spin current polarization is expressed by the constant $P$ with values ranging from $0 \leq P \leq 1$. In the following simulation we assumed a value of $P = 0.2$.

The study was carried out on a spin valve structure consisting of $Co_{90}Fe_{10}$ (20nm)/Cu (5nm)/$Ni_{80}Fe_{20}$ (5nm). The $Co_{90}Fe_{10}$ layer is considered as the fixed layer meaning that the magnetization is pinned in terms of the spin torque driven magnetization processes due to a larger thickness, $d$, and higher saturation magnetization, $M_s$, compared to the $Ni_{80}Fe_{20}$ free layer. The spin valve system is contacted with two Cu point contacts (40nm in diameter) with a centre to centre distance of 500 nm. The overall size of the simulation area was 1000 nm in diameter, with a discretization size of 4.5nm . The system is current driven, meaning that the changes in the relative orientation of the magnetization in the free and fixed layer are seen as voltage changes due to the GMR effect.

To verify the micromagnetic model against experiments we performed simulations with an external field of 740 mT applied at variable out of plane angles, ranging from 15° to 75° and a fixed current density of $J_e = 7.16 \times 10^{12}$ A/m² through one of the point contacts. All simulations are performed over a time period of 20 ns and with the new boundary method [20] a stable oscillation is achieved after 5 ns. It has been shown that by introducing artificial surface roughness spin waves are no more reflected at the boundary of the circular sample [20]. Therefore the stable oscillations modes are found and do not break down by destructive interferences. This technique is similar to other means of suppressing spin wave reflections such as the artificial increase of the Gilbert damping constant next to computational boundary [13, 21].

The results of the calculations are summarized and compared with experimental results in figure 1. It gives the frequency as a function of applied field angle and shows clearly a red shift in frequency as the out of plane angle is increased. This coincides with the experimental results (frequency and shift in frequency with field angle) published by Rippard [5]. The relative difference of between the calculated (this work) and the measured frequencies [5] is less than 0.04 over the entire range of field angles.

After the validation of the model on a single point contact, we studied the magnetization dynamics of a double point contact structure. In particular we investigated the magnetisation dynamics that lead to a phase locking behaviour and its dependence on different external field angles. Further the external field was applied in a way so that the projection of the field onto the film coincides with the line connecting the two point contacts. Starting from the same initial magnetisation state below each point contact (A and B), which is in plane and at an angle of 90° with respect to the reference layer, we applied a fixed current through point contact A ($J_{e,A} = 7.16 \times 10^{12}$ A/m$^2$) and a varying current through point contact B ($J_{e,B} = 5.57 \times 10^{12}$ A/m$^2$ to $J_{e,B} = 2.39 \times 10^{13}$ A/m$^2$). The current through contact B is increased in 1 mA increments while the one through contact A is fixed. For each current pair ($J_{e,A}$, $J_{e,B}$) the simulation was started from a well defined initial state as defined above. After an initial phase of chaotic magnetisation processes due to the chosen initial configuration, the system stabilizes and the magnetization vectors underneath each contact begin to oscillate uniformly with small disturbances due to the incoming spin wave front from the opposing contact. Both contacts emit spin waves with frequency and wavelength dependent on the current density. The spin wave from the neighbouring point contact disturbs the low-angle uniform precession of the magnetisation near the point contact. This spin wave interference leads not only to a change in the full width at half maximum of the signal in the frequency domain (FWHM) but also to a shift of the oscillating frequency, $f$, with respect to the oscillation frequency of the neighboring point contact The simulations show that these changes of FWHM and $f$ due to spin wave interference fully depend on the polarization factor $P$, the applied current $J_e$, and the centre to centre distance between the contacts $l$. Here we investigate the influence of a variable current, $J_{e,B}$, on the properties of the double point contact with fixed $P$ and $l$.

For an out of plane field angle of 15°, a red shift in the frequency of the oscillations near contact B, $f_B$, occurs when the current through contact B, $I_B$, is increased (see figure 2a). The frequency of contact A is constant at $f_A = 24.505$ GHz ($I_A = 9$ mA) and the one from contact B changes from $f_B = 25.320$ GHz to $f_B = 24.276$ GHz when the current is increased from $I_B = 7$ mA to $I_B = 30$ mA. The change in frequency with current is 1.053 GHz over 23 mA giving a tunability of 45 MHz/mA. Phase locking is reached after increasing the current in contact B to 21 mA which results in a locking current difference $\Delta I_{AB}$ of 12 mA.

We follow the experimental set up of [8] and changed the current through one contact with the current fixed in the other contact. Fig 2 shows the frequencies of the magnetization near point contact A and near point contact B. When the frequency as function of current does not change with current, this is a clear indication that the two contacts are phase locked. This condition is also seen by a change in the Fourier spectrum of the magnetization which shows a single peak only when phase locking occurs. The insets (0910 and 0922) in figure 2a and 2b show the Fourier spectrum (Lomb periodogram) for two current configurations. In inset 0910 the current through contact A is 9 mA and 10 mA through contact B. The Fourier analysis of the magnetization signal in the x-direction show two distinguishable peaks one at 24.49 GHz (contact A) and one at 25.039 GHz (contact B). The difference in frequency can be explained by the localized change in magnetization by the incoming spin waves from neighboring point-contact which also changes the local Oersted field distribution by a change in magneto resistivity. In the following we keep the current through contact A fixed and change the current through contact B. By increasing the current in contact B to 22 mA, a red shift in frequency occurs and the Fourier transform of the magnetization, see inset 0922, shows one peak with twice the power (the normalized power is the magnitude squared of the Lomb spectrum, divided by the data set size and the variance of the data series), which indicates phase-locking. The phase locking regime extends over a current range of 2mA, see figure 2a. A difference in locking currents, $\Delta I_{AB}$, is also observed experimentally [8]; phase locking occurs in a regime where the currents through the two contacts are different. Experimentally $\Delta I_{AB}$ is about 1 mA for a field angle of 75°. We assume that the locking-current difference is a result of spin wave interference between the point contacts. Spin waves emitted from one contact changes the oscillation behaviour of the other contact. Therefore the exact conditions below the contacts are not the same even if the geometry is symmetric.

By increasing the applied field angle to 30° two things can be observed: a red shift in frequency and a shift of the locking regime to higher currents (figure 2b). The tunability changes to 31 MHz/mA and $\Delta I_{AB}$ is increased to 14 mA. For an out of plane field angle of 45° the slope of the curve $f_B(I_B)$ becomes smaller, so that the frequency decreases with increasing current at a rate of 21 MHz/mA; the locking current difference $\Delta I_{AB}$ drops to 6 mA(figure 3a). For high out of plane field angles of 75°, the frequency of the magnetization oscillations near contact B first decreases and the one from A increases until they oscillate with the same frequency, see figure 3b. The phase locking regime persists until the current in contact B is increased by 1 mA, where after two well distinguished frequency are observed, and show a blue shift in frequency with current. The average increase in frequency with current is 179 MHz/mA. $\Delta I_{AB}$ is reduced to 1 mA. This value is similar to the locking-current difference found experimentally for a field angle of 75° [8].

We presented a finite element micromagnetic model that includes the asymmetric Slonzewski term and uses modified boundary conditions to minimize reflections from boundaries that is capable to simulate point contact spin torque oscillators.

Micromagnetic simulations were performed to find the conditions for mutual phase locking of two point contacts. Although the red shift (at low field angle) and blue shift (at high field angle) in frequency as function of current for single point contacts is supported by experimental data [5] there are no measurements on double point contacts on phase locking for low applied field angles.

We show that the phase locking in point contact geometries can be explained by the asymmetric Sloncewski spin torque term and that the locking current difference is a function of applied field angle, which has its minimum of 1m A at high angles (75 degree) and reaches a maximum of 12 mA at low angles (15 degree).

This work was supported by the European Communities programs IST STREP, under Contract No. IST-016939 TUNAMOS

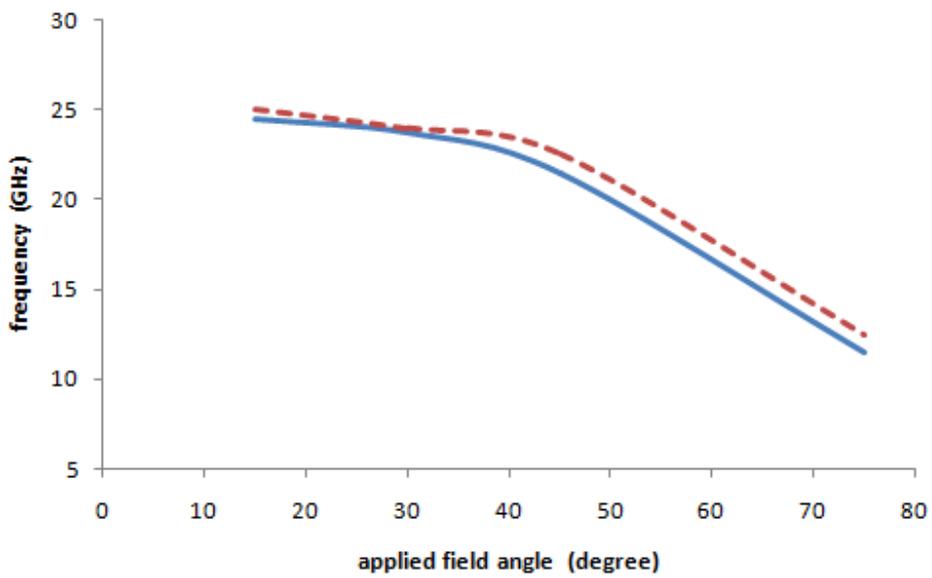

FIG. 1.(colour online) Frequency as function of the out of plane angle of the applied field ($|H_{ext}|$ = 740 mT) for a current ($I$ = 9 mA) through one point contact with a diameter of 40 nm. These results, micromagnetic simulations (bold line) agree with the experimental data (dotted line) reported in [5].

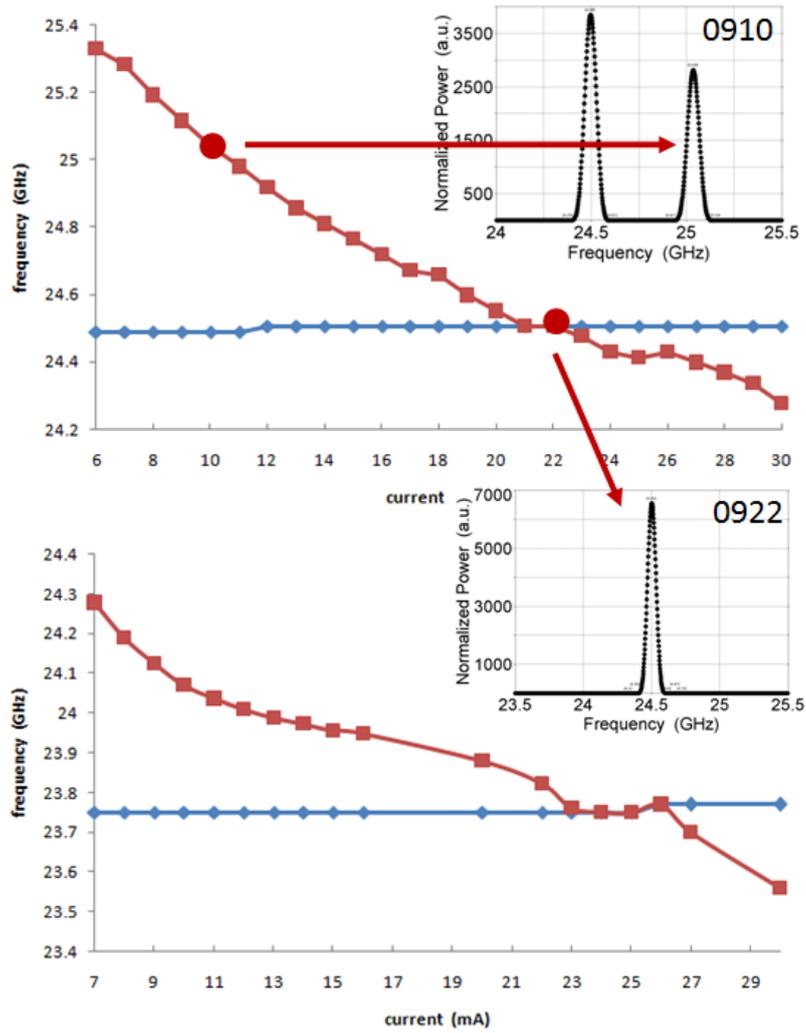

FIG. 2. .(colour online) (a) Double point contact, frequency of the magnetic oscillations near contact A (diamonds) and contact B (squares) as function of current for 15 ° out of plane field. The current through contact A is fixed at 9 mA and the current through contact B is varied from 6.5 mA to 30 mA. The flat region in curve B near 21 mA clearly indicates phase locking. (b) Frequency as function of current for 30° out of plane field. Phase locking occurs for $23\text{mA} \leq I_B \leq 26\text{mA}$, $I_A$ is fixed at 9 mA. The insets in (a) and (b) show the Fourier spectrum for the 15 ° case. In 0910 the current through contact A is 9 mA and 10 mA through contact B (showing two distinguishable peaks) and in 0922 the current through contact B is 22 mA and 9 mA through contact A (showing one peak with twice the power, which indicates phaselocking).

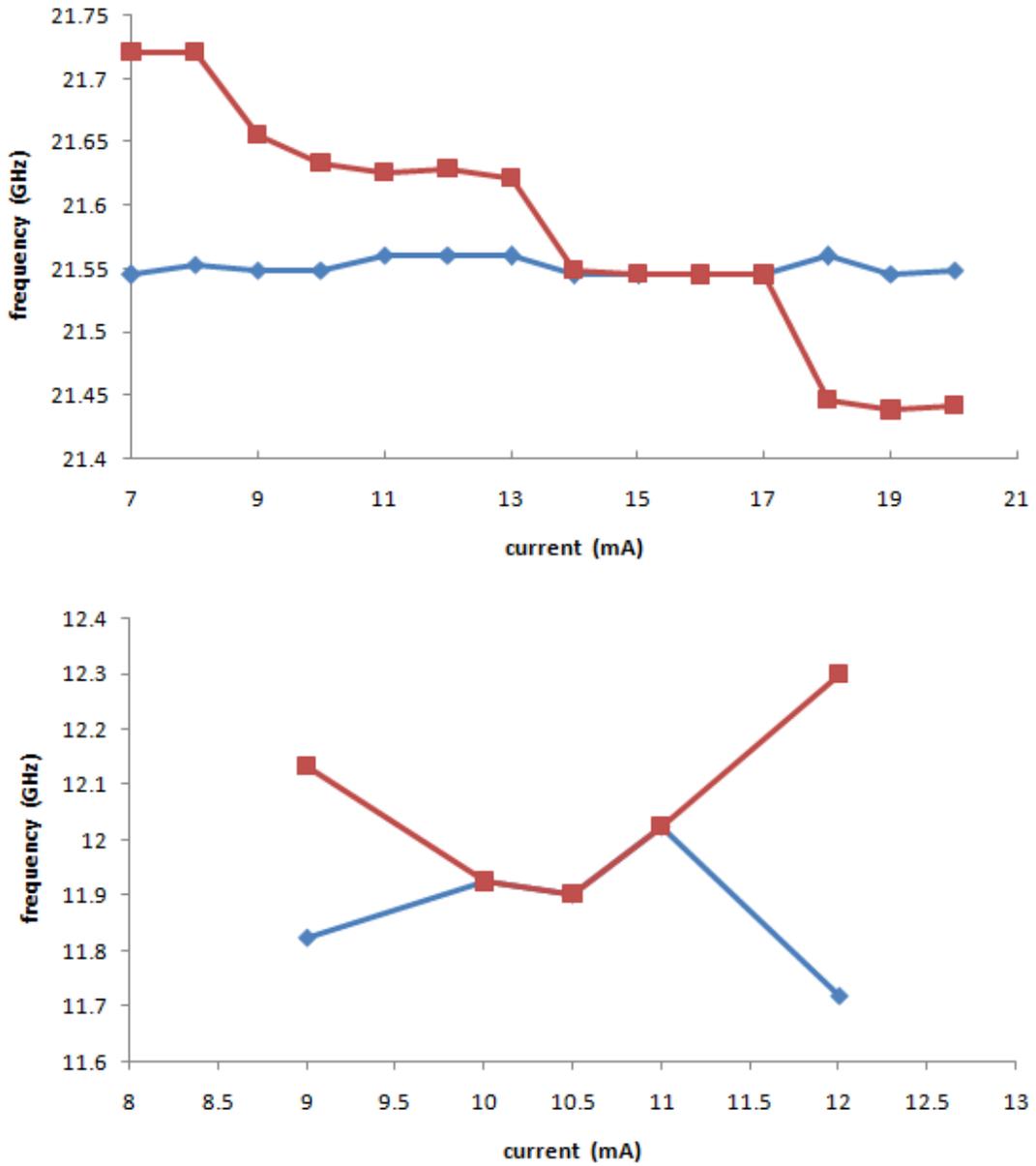

Fig. 3. .(colour online) (a)Double point contact, frequency of the magnetic oscillations near contact A (diamonds) and contact B (squares) as function of current for 45 ° out of plane field. The current through contact A is fixed at 9 mA and the current through contact B is varied from 7 mA to 20 mA. The flat region in curve B near 15 mA clearly indicates phase locking. (b) Frequency as function of current for 75 ° out of plane field. Phase locking occurs for $10\text{mA} \leq I_B \leq 11\text{mA}$, $I_A$ is fixed at 9 mA